# Latent Class Analysis of Digital Content Use Contexts in AI-Generated Synthetic Personas

Eunjeong Song · Sehee Hong
Department of Education, Korea University, Seoul, Republic of Korea

Eunjeong Song, ORCID iD: https://orcid.org/0000-0002-2302-3227
Sehee Hong, ORCID iD: https://orcid.org/0000-0001-5468-8398

## Abstract

AI-generated synthetic personas are increasingly used for content planning and virtual-user simulation, yet the digital content use contexts embedded in their narratives remain underexamined. Using all 1,000,000 records of NVIDIA's Nemotron-Personas-Korea, this study coded mentions of five engagement modes—viewing, discovering and sharing, interacting, reading, and listening—with an openly released Korean coding dictionary and applied latent class analysis with a three-step approach accounting for classification error. Four classes emerged: viewing-centered (20.8%), reading-centered (3.8%), listening-centered (2.1%), and low-mention (73.3%). Age showed the strongest association: the odds of the viewing-centered versus low-mention class were multiplied by 0.15 per 10-year increase, while mentions of any non-viewing mode declined from 49.4% among those under 30 to 2.3% among those aged 70 and over. Sex differences in predicted class probabilities were below 1 percentage point, and provincial differences reached at most 5.0 percentage points. These classes describe narrative configurations, not consumer segments, and provide a basis for assessing diversity in synthetic persona use contexts.

## Introduction

Generative artificial intelligence (AI) now produces user profiles and situational narratives automatically, extending the reach of content planning, service ideation, review of recommendation results, and virtual-user simulation (Batzner et al., 2025; De Paoli, 2026; Salminen et al., 2024). Whereas traditional personas consisted of a small number of designed archetypes representing typical users, AI-generated synthetic personas can be produced at scale as narrative data conditioned on diverse demographic attributes. Such data can be used to explore a range of use situations at the early planning stage, where research on actual users is difficult.

AI-generated synthetic personas, however, are not observational data on actual users. A generative model composes a plausible lifestyle and plausible content-use scenes from conditions such as age, sex, occupation, and region, and in doing so it may repeatedly couple particular demographic attributes with particular platforms or content, or reduce the use contexts of some groups to stereotypes (Batzner et al., 2025; Haxvig et al., 2025; Salminen et al., 2024). Before synthetic personas are used as proxies for content consumers or market segments, therefore, it is necessary to verify empirically what digital content use contexts the data actually contain. Prior studies have addressed the use of AI-generated synthetic personas for ideation (De Paoli, 2026), the composition and bias of generated narratives (Haxvig et al., 2025; Salminen et al., 2024), and representativeness and transparency of reporting (Batzner et al., 2025). The present study extends this discussion to an openly released, large-scale Korean-language data set by examining which engagement modes are mentioned together across the entire data set and how their configuration relates to demographic attributes.

Nemotron-Personas-Korea (hereafter NPK), released by NVIDIA, is a Korean-language synthetic persona data set of 1,000,000 records that pairs structured attributes such as age, sex, and province of residence (one of the 17 si/do, the first-level administrative divisions of Korea) with seven persona narratives, ranging from a professional persona to a general persona, and with narrative fields on cultural background, skills, hobbies, and career goals (H. Kim et al., 2026). This structure makes it possible to analyze how the generative model links demographic attributes to content-related narratives. Nevertheless, the combinations in which digital content touchpoints appear in NPK narratives, and how those combinations relate to demographic attributes, have not yet been sufficiently examined.

Research on media use has characterized users not by the amount of use of any single medium but by the combination of media and platforms used together, termed the media repertoire (Hasebrink & Popp, 2006; Lee & Park, 2020; Taneja et al., 2012). This study applies that perspective to the NPK narrative data and operationalizes digital content use context as five engagement modes: viewing, discovering and sharing, interacting, reading, and listening. In this study, engagement mode refers to these five indicators, and use-context type refers to a latent class defined by how the engagement modes combine within a single persona; the rationale for the indicators and the coding criteria are presented in the Method section.

Moreover, the mention rates of individual engagement modes alone reveal little about how the modes combine within a persona. Latent class analysis (LCA) estimates unobserved subtypes from the response patterns of multiple categorical indicators and summarizes the co-occurrence structure through class-specific response probabilities (Collins & Lanza, 2010; Nylund et al., 2007); with a three-step approach that accounts for classification error, the associations of age, sex, and region with class membership can be analyzed while the measurement structure of the latent classes is held fixed (Asparouhov & Muthén, 2014; Vermunt, 2010).

The purpose of this study is to classify the co-occurrence of the five digital content engagement modes in NPK into latent classes and to identify how each type relates to age, sex, and province of residence. In doing so, we aim to show what digital content use contexts AI-generated synthetic personas construct and to provide an empirical basis for interpreting synthetic personas in content planning and user simulation. The research questions are as follows.

*Research Question 1:* What latent classes are formed by the co-occurrence of the five digital content engagement modes in the AI-generated synthetic personas of NPK?

*Research Question 2:* How are age, sex, and province of residence related to the probability of membership in the latent classes of digital content use context?

# Method

## Data

This study analyzed all 1,000,000 records of the base release of NPK version 1.0, which NVIDIA released on Hugging Face under a CC BY 4.0 license (H. Kim et al., 2026). Because the full data set is analyzed, the target of statistical inference is the process that generated NPK, and the 1,000,000 records are regarded as a sample from that generative distribution. For reproducibility, only the files registered under commit 03d5650 were used. NPK consists of structured demographic and geographic variables and several Korean-language narrative fields; among the structured variables, we used age ('age'), sex ('sex'), and province of residence ('province').

For coding the digital content engagement modes, all text fields were used, including the professional, sports, arts, travel, culinary, family, and general persona fields and the fields on cultural background, skills and expertise, hobbies and interests, and career goals. Because each indicator registers whether a mode is mentioned anywhere in a record, a mode repeated across several fields was counted once. No record was entirely without text, so all 1,000,000 records were included in the analysis.

This study used only openly released, fully synthetic data that contain no responses from real individuals and no personally identifiable information, and therefore does not constitute human subjects research.

## Digital Content Engagement-Mode Indicators

### *Theoretical Rationale for the Indicators*

The digital content engagement-mode indicators were organized not by platform company or content genre but by the persona's primary content engagement behavior—that is, from the perspective of media as practice, which asks what people do with media (Couldry, 2004). In terms of the user's role, we distinguished reception, in which content is taken in; participation, in which the user's manipulation and feedback become part of how the content unfolds (Steuer, 1992); and mediation, in which content is explored and circulated to others (Jenkins et al., 2013). Reception was further divided by semiotic modality into viewing (watching video content), reading (reading text- and image-based digital publications), and listening (listening to music and voice content). Participation corresponds to the interacting (gaming) indicator and mediation to the discovering and sharing indicator; in this study, interacting refers to game-type interaction. This distinction also parallels the Korea Media Panel Survey, which treats media activities such as video, music, and gaming, as well as the use of social network services (SNS) and over-the-top (OTT) services, as separate categories (Y. H. Kim, 2025). Because engagement modes are not mutually exclusive person types, several modes may appear in one persona. E-commerce, delivery, financial, map, transit, and reservation services are digital platforms, but they serve transactions or utility functions rather than content use and were excluded from the main indicators. The concept and coding criteria for each indicator are presented in Table 1.

**Table 1**

*Concepts and Coding Criteria for the Digital Content Engagement-Mode Indicators*

| Indicator | Primary content engagement behavior | Example expressions[a] | Boundary-case handling |
|---|---|---|---|
| Viewing | Watching online video, short-form video, live streaming, and OTT content | YouTube ('유튜브') videos, Netflix ('넷플릭스'), TVING ('티빙'), Wavve ('웨이브'), Watcha ('왓챠'), online streaming | General TV viewing is excluded; included only when a digital medium is specified |
| Discovering & Sharing | Discovering and sharing content, and related communication, on social media and online communities | Instagram ('인스타그램'), TikTok ('틱톡'), Facebook ('페이스북'), online communities, KakaoTalk open chat ('카카오톡 오픈채팅'), Naver Cafe ('네이버 카페') | Plain messenger contact is excluded; included when a content discovery or sharing context is present |
| Interacting (Gaming) | Exchanging controls, choices, and feedback in digital games | Online games, mobile games, console games, Steam ('스팀'), named game titles | Watching e-sports is coded as Viewing; playing games is coded as Interacting (Gaming) |
| Reading | Reading digital publications such as webtoons, web novels, and e-books | Naver Webtoon ('네이버웹툰'), KakaoPage ('카카오페이지'), RIDI ('리디'), Millie's Library ('밀리의서재'), web novels, e-books | Mentions of print books only are excluded |
| Listening | Listening to sound-based content such as digital music, podcasts, and audiobooks | Melon ('멜론'), Genie ('지니'), Spotify ('스포티파이'), YouTube Music ('유튜브뮤직'), podcasts, audiobooks | Offline concert attendance and generic singing activities are excluded |

*Note.* OTT = over-the-top. Service names are illustrative; rules for Korean and English variants, abbreviations, spacing variants, and ambiguous expressions are specified in the openly released coding dictionary.
[a] Korean strings in single quotation marks are the verbatim dictionary terms applied to the Korean-language text.

### *Dictionary Construction and Coding Rules*

The coding dictionary was initially assembled by combining the names of major domestic and international services, expressions observed repeatedly in a preliminary review of NPK, and generic terms designating each engagement mode. Text was normalized (Unicode normalization and case-folding of English letters), and spacing, hyphenation, Korean and English script variants, abbreviations, and word boundaries were handled consistently. Where a single platform provides several functions, the described behavior took precedence over the platform name: watching videos on YouTube was coded as viewing, whereas sharing content through a YouTube channel or comments was coded as discovering and sharing. A service name without a stated behavior was coded according to the platform's principal content function, so that '유튜브' (YouTube) alone was coded as viewing, whereas a service name that itself specifies the engagement mode, such as '유튜브 뮤직' (YouTube Music), was coded as listening. When different behaviors were stated in a single sentence, each corresponding indicator was coded 1.

Ambiguous terms that overlap with everyday vocabulary—such as '멜론' (Melon), '지니' (Genie), '웨이브' (Wavve), and '스팀' (Steam)—were coded only when an expression related to the corresponding engagement mode (a collocate) appeared in the same sentence (hereafter, the accompanying-expression condition). For example, '멜론' was coded as listening only when it appeared in the same sentence as a listening-related expression such as music, song, listening, streaming, or playlist, and never in a fruit or food context; all accompanying-expression conditions are specified in the coding dictionary. Expressions that name no service, such as '영상 시청' (watching videos) and '음악 감상' (listening to music), do not make digital use evident and were conservatively excluded from the main analysis. In contrast, generic expressions in which the digital medium is unambiguous, such as 'OTT 시청' (OTT viewing), '온라인게임' (online game), '웹툰 앱' (webtoon app), and '팟캐스트' (podcast), were included.

After the dictionary was constructed, high-frequency expressions across the full text were tabulated to check for missing service names and generic expressions, and any omissions were added to the dictionary. The complete coding dictionary and the preprocessing and coding rules are provided as open materials.

Because coding applies a fixed dictionary and fixed rules to the text, identical input reproduces identical output. Inter-coder reliability, which concerns the judgment stage of human coders, is therefore not the relevant criterion; we instead evaluated coding quality by the sensitivity of the results to the choice of rules and by the precision with which the rules were applied. Sensitivity was evaluated by repeating the full coding with two dictionary variants—a reduced dictionary that excludes ambiguous and generic expressions and contains only unambiguous proper names, and an extended dictionary that additionally includes generic expressions without service names—and re-estimating the unconditional model (which contains no covariates) under each variant, to assess whether the decision on the number of classes and the class-specific profiles depended on the composition of the dictionary. Precision was evaluated through an audit: with the match–mode pairs produced by the final coding rules as the population, 100 pairs per engagement mode (500 in total) were drawn at random and judged correct or incorrect by one author. Overall precision is the weighted mean of the per-mode precisions, with weights equal to each mode's share of the population matches; the method for computing intervals and the judgment criteria are presented in Table A2 in the Appendix and in the open materials.

**Latent Class Analysis**

Latent class models were estimated under the assumption of local independence, that is, that the five binary indicators are mutually independent within each latent class (Collins & Lanza, 2010). LCA estimates, for each class, the probability of responding 1 on each indicator given membership in that class; because the indicators in this study register whether a mode is mentioned in the narrative, we denote this probability as the conditional mention probability (hereafter, mention probability). The class-specific mention probabilities served as the principal basis for interpreting the digital content use context of each class.

Because the observed information provided by five binary indicators consists of 31 independent response-pattern proportions, a six-class model with 35 free parameters is not identified, and a five-class model with 29 free parameters is close to saturated. We therefore compared the one- to four-class models as the main competing models and examined the five-class model only as an auxiliary model, to check whether it added a new, interpretable use context.

Text preprocessing, dictionary-based coding, and the generation and execution of Mplus input files were performed in Python 3.12.3, and all latent class models and the R3STEP analysis were run in Mplus 9.1 (Muthén & Muthén, 2026). To reduce the computational burden, the unconditional models and the likelihood ratio tests were estimated on frequency-weighted data consisting of the 32 response patterns, which summarize the observed data exactly; the covariate analysis used the record-level data. The number of latent classes was decided by jointly considering normal termination and repeated replication of the best log-likelihood; the adjusted Lo-Mendell-Rubin (LMR) likelihood ratio test (Lo et al., 2001); the information criteria, namely the Akaike information criterion (AIC), the Bayesian information criterion (BIC), and the sample-size adjusted BIC (aBIC); classification quality as evaluated by entropy and average posterior class-membership probabilities; the proportion of the smallest class; and the interpretability and parsimony of the mention-probability profiles (Collins & Lanza, 2010; Nylund et al., 2007).

The parametric bootstrapped likelihood ratio test (BLRT; McLachlan & Peel, 2000) was not computed because Mplus does not support TECH14 for frequency-weighted mixture models. For the adjusted LMR test, we verified that the $(k-1)$-class solution reproduced the best log-likelihood for that number of classes, and comparisons in which it was not reproduced were recomputed with an increased number of random starts for the $(k-1)$-class model.

Because, when the full data set is analyzed, likelihood ratio tests and information criteria can favor an additional class even when the improvement is negligible, we fixed the following rule before the analysis: the number of classes would not be increased on the basis of sample size or *p* values alone, and the primary criterion would be whether an additional class presents a new combination of engagement modes, merely subdivides an existing class by overall mention level, or merely isolates a very small number of observed patterns. Entropy was used only as an index describing classification quality, not as a principal criterion for deciding the number of classes (Collins & Lanza, 2010), and values of .4, .6, and .8 were interpreted as benchmarks for low, medium, and high levels of classification, respectively (Clark & Muthén, 2009). A smallest class below 5% prompted scrutiny of whether the class had arisen by chance (Berlin et al., 2014), but whether the class showed a distinct combination of modes took priority over mechanical application of the threshold.

To reduce the risk of convergence to local maxima, the preliminary analyses used 500 random starts (125 final-stage optimizations), and for the final candidate models the number of starts was doubled to 1,000 (250) to confirm replication of the best log-likelihood. The local independence assumption was checked with the standardized residuals of the bivariate contingency tables.

**Covariates and Latent Class Membership**

After the final class model was fixed, age, sex, and province of residence were entered simultaneously as covariates of latent class membership using R3STEP, the automated implementation in Mplus of the three-step approach that accounts for classification error (Asparouhov & Muthén, 2014; Vermunt, 2010). This method treats the covariates as auxiliary variables: the latent classes established in the Step 1 unconditional model are retained and classification error is held fixed as logits while the covariate effects are estimated, so that the formation of the classes is unaffected by the covariates. The method, however, presupposes that the mention probabilities of the indicators within a class do not vary with the covariates (Asparouhov & Muthén, 2014); this assumption was checked by contrasting the observed mention rates by age band with the model-implied mention rates.

For ease of interpretation, age was rescaled to 10-year units; sex used the categories of the source data (female = 1); and province of residence was entered as 16 dummy variables with Seoul as the reference province. No record was missing on age, sex, or province. The reference class for the odds ratios is the low-mention class. The results for age and sex are reported as odds ratios (*OR*s) relative to the reference class, 95% confidence intervals (CIs), and average differences in predicted probability; the predicted-probability difference is the change in a record's predicted probability when its age is increased by 10 years (for sex, when female is set to 1 rather than 0) with its other covariates held at their observed values, averaged over all records. The province-specific predicted probabilities are obtained by setting the province of all records to the given province and averaging over the distribution of age and sex; the range of predicted probabilities for each class and the provinces with high and low probabilities are summarized in the text, and the full results are presented in the Appendix and the open materials. Predicted probabilities and their differences do not depend on the choice of reference class. Because the analysis covers the full data set, interpretation focused on the magnitude of odds ratios and predicted-probability differences rather than on $p$ values.

# Results

## Mention Rates by Engagement Mode

As shown in Table 2, the mention rates of the digital content engagement-mode indicators ranged from 3.71% to 41.50%, and the φ correlations between indicators ranged from −.07 to .09, so no pair of indicators was highly redundant. The type of expression that triggered a positive code differed markedly across engagement modes: 97.8% of viewing codes rested on unambiguous proper service names, whereas only 0.7% of listening codes did, reflecting a writing convention in which listening is described mainly through generic medium terms such as podcasts and music streaming. This asymmetry limits what the dictionary-variant sensitivity analysis can show and is therefore interpreted together with that analysis in the Latent Classes of Digital Content Use Context section. The detailed composition of evidence by engagement mode, and the number of ambiguous candidate matches suppressed by the accompanying-expression condition, are given in the open materials. In the precision audit, match-level precision by engagement mode ranged from .920 to 1.000, and overall precision weighted by the number of final emitted matches was .984 (95% interval [.958, .991]; Table A2). Errors were concentrated in discovering and sharing (8 cases): in 6, '유튜브' (YouTube) had been reclassified as discovering and sharing because '채널' (channel) or '공유' (sharing) appeared in the same sentence, and in 2, SNS and KakaoTalk were mentioned in contexts other than content use (profile design work and setup assistance).

**Table 2**

*Mention Rates of Digital Content Engagement Modes*

| Engagement mode | Records | Share of total (%) |
|---|---|---|
| *Individual modes* | | |
| Viewing | 415,021 | 41.50 |
| Discovering & Sharing | 69,879 | 6.99 |
| Interacting (Gaming) | 55,445 | 5.54 |
| Reading | 41,102 | 4.11 |
| Listening | 37,146 | 3.71 |
| *Overall (record level)* | | |
| At least one mode mentioned | 517,548 | 51.75 |

| Engagement mode | Records | Share of total (%) |
|---|---|---|
| Two or more modes mentioned | 90,177 | 9.02 |
| No mode mentioned | 482,452 | 48.25 |

*Note.* The denominator is the 1,000,000 records included in the analysis.

Viewing was by far the most commonly mentioned mode (41.5%), whereas the other four modes were rare (3.7% to 7.0%); 9.0% of records mentioned two or more modes, and 48.3% mentioned none.

## Number of Latent Classes

Table 3 presents the model comparison for deciding the number of classes.

**Table 3**

*Model Comparison for Determining the Number of Latent Classes*

| Classes | Log-likelihood | AIC | BIC | aBIC | Adjusted LMR *p* | Entropy | Smallest class (%) |
|---|---|---|---|---|---|---|---|
| 1 | −1,476,401.73 | 2,952,813.46 | 2,952,872.54 | 2,952,856.65 | – | – | – |
| 2 | −1,468,659.75 | 2,937,341.51 | 2,937,471.48 | 2,937,436.52 | < .001 | 1.000 | 41.5 |
| 3 | −1,464,392.04 | 2,928,818.09 | 2,929,018.95 | 2,928,964.92 | < .001 | .413 | 22.3 |
| 4 | −1,463,993.97 | 2,928,033.95 | 2,928,305.71 | 2,928,232.61 | < .001 | .630 | 2.1 |
| 5 | −1,463,978.56 | 2,928,015.12 | 2,928,357.77 | 2,928,265.61 | < .001 | .477 | 0.8 |

*Note.* AIC = Akaike information criterion; BIC = Bayesian information criterion; aBIC = sample-size adjusted BIC; LMR = Lo-Mendell-Rubin. The bootstrapped likelihood ratio test was not computed for the frequency-weighted data; the five-class model was examined only as an auxiliary model.

All models from one to five classes terminated normally, and the best log-likelihood was replicated repeatedly even with the doubled number of random starts (nine times for the four-class model). In the auxiliary five-class solution, however, the best log-likelihood was observed only once, so that solution may be unstable. In models with two or more classes, the thresholds of indicators whose mention probability in a particular class approached 0 or 1 were fixed at extreme values (e.g., viewing in the reading-centered class and listening in the listening-centered class in Table 4), which is expected for sparse responses within a class. BIC and aBIC were both lowest at four classes, whereas AIC was lowest at five; when the information criteria disagreed, precedence was given to BIC, which performs relatively well in simulation studies on deciding the number of classes (Nylund et al., 2007). The adjusted LMR likelihood ratio test yielded $p < .001$ for every adjacent comparison and thus, as anticipated in the Latent Class Analysis section, contributed nothing to discriminating among class solutions in the full data set. The five-class model increased BIC and aBIC relative to the four-class model, its smallest class was only 0.8%, and, rather than presenting a new combination of engagement modes, it merely rearranged the solution by splitting the low-mention class into two classes differing in the level of viewing mentions and shrinking the listening- and reading-centered classes. In addition, the likelihood ratio chi-square against the 32 observed response patterns was $\chi^2(14) = 840.99$ for the three-class model, $\chi^2(8) = 44.86$ for the four-class model, and $\chi^2(2) = 14.03$ for the five-class model, and the sum of absolute differences between observed and expected frequencies was 8,602, 675, and 394 of the 1,000,000 records, respectively; the four-class model thus reproduced the observed patterns almost exactly. Weighing together response-pattern fit, BIC and aBIC, the stability of the solution, class size, and the substantive interpretability of an additional class, we selected the four-class model as the final model.

The entropy of the final model was .630, corresponding to a medium level of classification (Clark & Muthén, 2009), and the average posterior class-membership probabilities presented in Table 4 ranged from .68 to .83. This classification uncertainty limits record-level class assignment, which is why the covariate analysis used the three-step approach that accounts for classification error. The smallest class (listening-centered) constituted 2.1% of the total. Although this proportion falls below the 5% benchmark, the class was retained because it corresponds to approximately 21,000 of the 1,000,000 records and, as shown in Table 4, has a distinct profile in which the listening mention probability converges to 1.00 while the remaining four indicators are all at or below .10, so that listening alone separates it clearly. In the final model, the maximum bivariate standardized residual was 0.68, providing no substantive evidence of a violation of local independence.

**Latent Classes of Digital Content Use Context**

Class names were assigned after inspecting the class-specific mention-probability profiles, on the basis of the prominent engagement mode and the overall level of mention. Table 4 and Figure 1 present the conditional mention probabilities by class.

**Table 4**

*Conditional Mention Probabilities and Class Proportions*

| Engagement mode | Viewing-Centered | Reading-Centered | Listening-Centered | Low-Mention |
|---|---|---|---|---|
| Viewing | .658 | .000 | .100 | .377 |
| Discovering & Sharing | .261 | .098 | .056 | .014 |
| Interacting (Gaming) | .124 | .125 | .056 | .032 |
| Reading | .076 | .505 | .050 | .007 |
| Listening | .021 | .023 | 1.000 | .014 |
| Class proportion (%) | 20.8 | 3.8 | 2.1 | 73.3 |
| Average posterior probability | .756 | .689 | .682 | .826 |

*Note.* Average posterior probabilities are the diagonal entries of the classification table, that is, the mean posterior probability of the assigned class among records whose most likely class is that class. Classes are ordered as Viewing-Centered, Reading-Centered, Listening-Centered, and Low-Mention throughout.

**Figure 1**

*Conditional Mention Probabilities of Digital Content Engagement Modes by Latent Class*

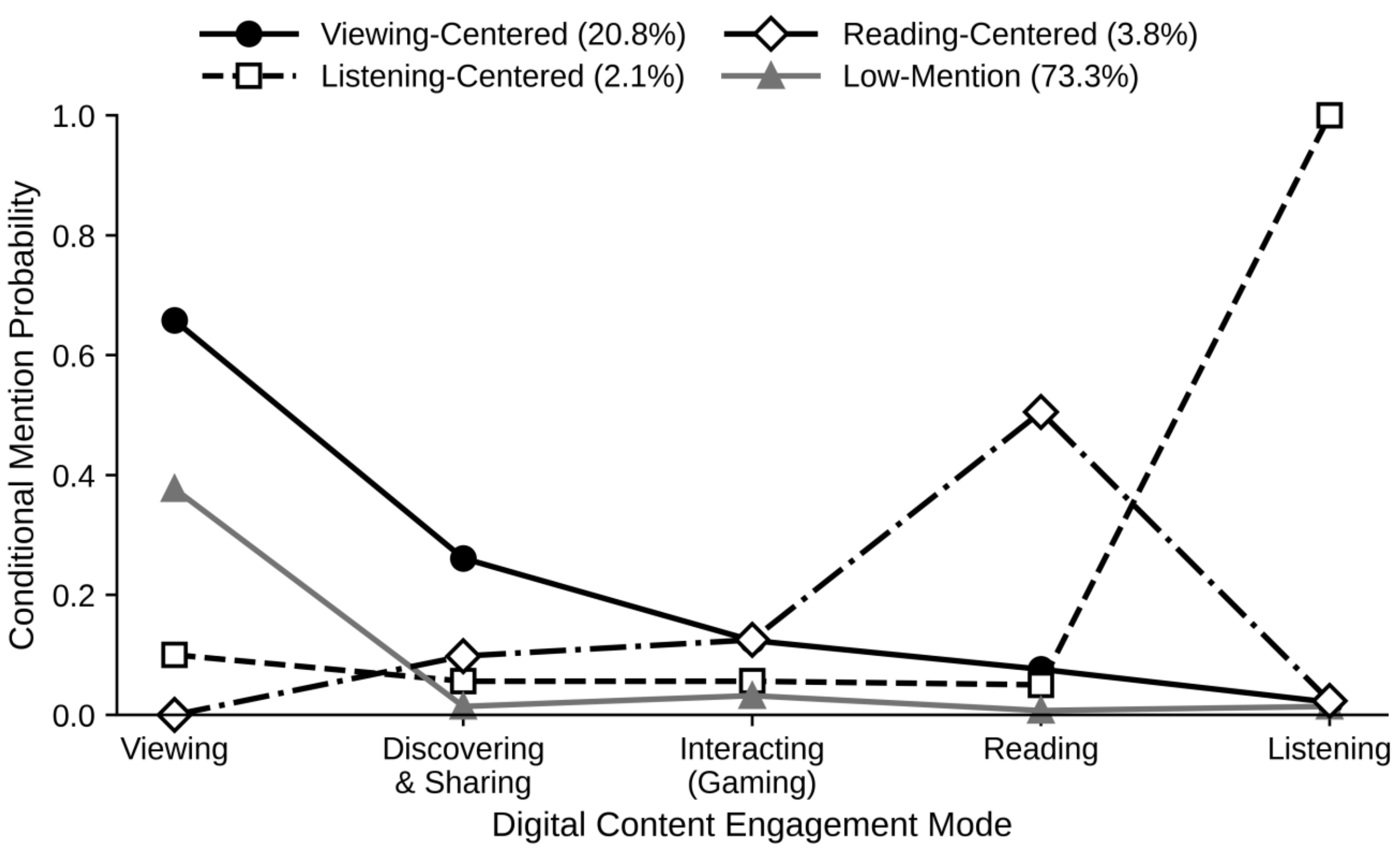

*Note.* Class proportions are given in parentheses in the legend.

The viewing-centered class (20.8%) combined a high viewing mention probability (.66) with an accompanying probability of discovering and sharing (.26). The reading-centered class (3.8%) was distinguished by a reading mention probability of .51 together with a viewing mention probability estimated at .00 (a boundary estimate); it thus represents narratives in which reading is described without viewing. The listening-centered class (2.1%) showed a pronounced single-mode profile: the listening mention probability converged to 1.00 (a boundary estimate) while all other indicators remained at or below .10. The low-mention class (73.3%) mentioned viewing only intermittently (.38), with all remaining indicators below .04.

As Figure 1 shows, the viewing-centered and low-mention classes are distinguished not by whether viewing is mentioned but by whether other engagement modes, particularly discovering and sharing (.26 vs. .01), accompany it. Weighted by the posterior class-membership probabilities of the final model, the low-mention class consisted of 58.2% no-mention patterns, 35.2% viewing-only patterns, and 6.7% other patterns. The posterior-weighted median total narrative length of the low-mention class was 1,453 characters, close to that of the other three classes (1,465–1,490 characters). The low-mention class was therefore interpreted not as a group of content non-users but as a narrative type in which mentions other than viewing were captured relatively infrequently under the coding criteria of this study. The relative ordering of mention probabilities across classes varied by engagement mode, so the four classes were distinguished by different combinations of engagement modes. Accordingly, the final classes were interpreted as narrative types representing distinct use contexts.

In the sensitivity analysis of dictionary variants, the information criteria under the extended dictionary continued to decrease through five classes, but the four-class solution reproduced the same type structure as the main analysis: viewing-centered (21.0%), reading-centered (6.2%), listening-centered (4.9%), and low-mention (67.9%). The reduced dictionary, which retains only proper names, by construction all but removes listening, for which proper-name evidence is sparse (mention rate 0.03%); it was therefore expected that the listening-centered class would disappear and that BIC and aBIC would be minimized at three classes under the reduced dictionary. We accordingly read the reduced-dictionary results as a robustness check on the four-indicator structure excluding listening.

**Associations of Age, Sex, and Province With Latent Class Membership**

Because the Step 1 solution of R3STEP reproduced the best log-likelihood and the class profiles of the final unconditional model, the results below refer to the latent classes defined in the preceding section. The results for age and sex are presented in Table 5.

**Table 5**

*Associations of Age and Sex With Latent Class Membership*

| Class contrast | Covariate | *OR* | 95% CI | Probability difference |
|---|---|---|---|---|
| Viewing-Centered vs. Low-Mention | Age (+10 years) | 0.149 | [0.145, 0.152] | −.1108 |
| | Female | 1.102 | [1.063, 1.143] | .0037 |
| Reading-Centered vs. Low-Mention | Age (+10 years) | 0.186 | [0.182, 0.191] | −.0193 |
| | Female | 1.370 | [1.315, 1.428] | .0095 |
| Listening-Centered vs. Low-Mention | Age (+10 years) | 0.410 | [0.404, 0.416] | −.0055 |
| | Female | 0.797 | [0.769, 0.827] | −.0059 |

*Note. OR* = odds ratio; CI = confidence interval. Odds ratios and 95% CIs are from the R3STEP analysis with Low-Mention as the reference class (see the Covariates and Latent Class Membership section). Probability differences are average changes in the focal-class probability for a 10-year increase in age (or female = 1 vs. 0) across all records and are invariant to the choice of reference class.

Each 10-year increase in age multiplied the odds of the viewing-centered class relative to the low-mention class by 0.15, 95% CI [0.145, 0.152], and the odds of the reading-centered (*OR* = 0.19) and listening-centered (*OR* = 0.41) classes also fell consistently. For sex, the clearest association was a higher probability of the reading-centered class among female personas (*OR* = 1.37); the listening-centered class leaned male (female *OR* = 0.80), and the viewing-centered class leaned slightly female (*OR* = 1.10). Judged by the magnitude of the predicted-probability differences, the association with age was substantial (−11.1 percentage points per 10 years for the viewing-centered class), whereas the association with sex was limited (below 1 percentage point in absolute value).

Province-specific predicted probabilities, averaged over the distribution of age and sex in the full data, ranged from 0.4% to 5.4% for the listening-centered class, a maximum difference of 5.0 percentage points. This class was most probable in Sejong (5.4%), Busan (5.2%), and Daegu (5.0%) and least probable in Gangwon (0.4%) and Seoul (0.6%). Among the other classes, the viewing-centered class was most probable in Gyeonggi (22.9%) and Incheon (21.1%), and the reading-centered class was most probable in Sejong (8.3%). Class-specific predicted probabilities for all 17 provinces are presented in Table A1 in the Appendix.

Figure 2 presents the class-specific predicted probabilities by age (Panel A, 20–90 years) and by sex (Panel B), computed from the same model with the other covariates averaged over the distribution of all records. Because age enters the between-class log-odds linearly, its effect is nonlinear on the probability scale: the decline of the viewing-centered class is steepest between ages 30 and 40 and flattens after age 50. The value of −11.1 percentage points in Table 5 is the 10-year change along this curve averaged over the age distribution of all records.

**Figure 2**

*Predicted Probabilities of Latent Class Membership by Age and Sex*

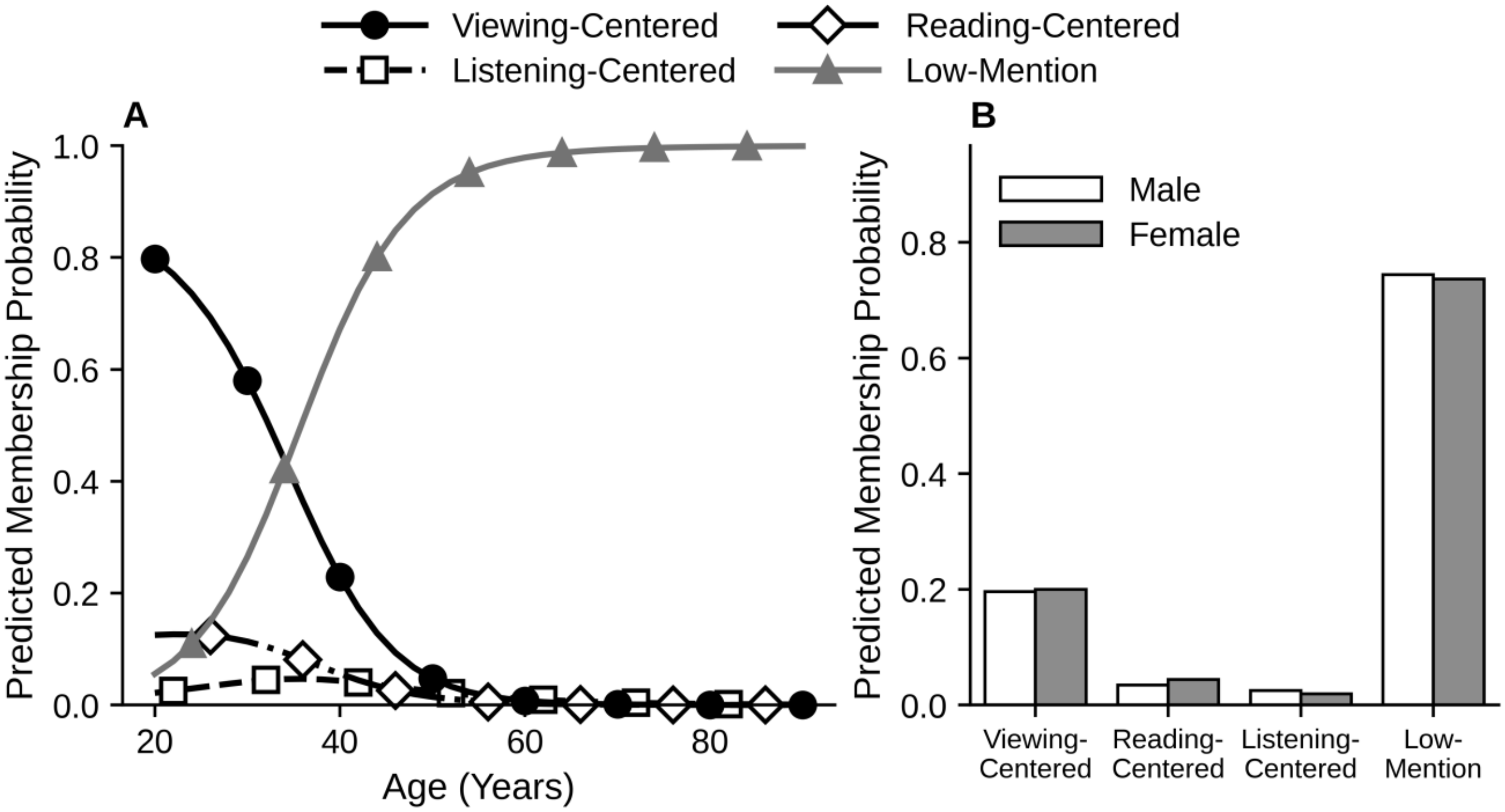

*Note.* Panel A varies age over 20–90 years; Panel B contrasts male and female personas. In both panels, the other covariates are averaged over the distribution of all records.

Table 6 presents the observed mention rates by age band, computed independently of the model. The proportion of records mentioning at least one mode other than viewing fell steadily across age bands, from 49.4% under 30 to 2.3% at 70 and over. Conversely, among records mentioning any mode, the share whose only mention was viewing rose steadily, from 35.4% under 30 to 90.3% at 70 and over. In other words, the content mentions captured for older personas were concentrated in viewing.

**Table 6**

*Observed Mention Rates of Engagement Modes by Age Band*

| **Age band** | ***N*** | **Viewing** | **Discovering & Sharing** | **Interacting (Gaming)** | **Reading** | **Listening** | **Any mode other than viewing** | **Viewing only, among records with any mention** |
|---|---|---|---|---|---|---|---|---|
| Under 30 | 145,502 | 52.1 | 22.3 | 18.3 | 14.4 | 4.9 | 49.4 | 35.4 |
| 30–39 | 149,659 | 45.2 | 9.1 | 11.7 | 7.6 | 6.3 | 31.0 | 52.1 |
| 40–49 | 175,647 | 43.0 | 5.4 | 4.8 | 3.4 | 6.0 | 18.3 | 66.8 |
| 50–59 | 198,750 | 45.2 | 3.5 | 1.0 | 1.0 | 3.1 | 8.3 | 83.4 |
| 60–69 | 178,738 | 40.4 | 2.9 | 0.4 | 0.4 | 1.6 | 5.2 | 87.9 |
| 70 and over | 151,704 | 22.4 | 1.3 | 0.1 | 0.2 | 0.7 | 2.3 | 90.3 |

*Note.* Values are percentages of records in each age band, computed directly from the coded indicators without reference to the latent class model.

The model-implied mention rates by age band under the measurement-structure assumption (see the Covariates and Latent Class Membership section) agreed in direction with the observed values but diverged for interacting among those under 30 (observed 18.3% vs. implied 11.0%) and for viewing among those aged 70 and over (22.4% vs. 37.6%). The age odds ratios in Table 5 should therefore be read as summaries under a fixed measurement structure, and the observed mention rates in Table 6 are presented alongside them to show how narrative composition changes with age.

Taken together, NPK generates digital content use contexts by attaching the full range of the five engagement modes to younger personas, skewing the captured mentions of older personas toward viewing, and concentrating listening and reading touchpoints in a subset of provinces.

# Discussion

## Main Findings

Analysis of the co-occurrence of viewing, discovering and sharing, interacting, reading, and listening in the narrative portions of the 1,000,000 NPK records yielded four latent classes, distinguished by the engagement mode central to the narrative (viewing, reading, or listening) and by whether content was mentioned at all. Because only 9.0% of records mentioned two or more modes, these classes are better understood as narrative types in which one mode is prominently mentioned, or in which mentions are generally sparse, than as use repertoires that actively combine multiple media. This shows that AI-generated synthetic personas do not merely list individual platforms independently but repeatedly construct particular combinations of digital content engagement behaviors.

Age, sex, and province of residence were all related to the probability of class membership, but not equally. Age stood out: each 10-year increase multiplied the odds of the viewing-centered class (relative to the low-mention class) by 0.15, and in the observed data too, the content mentions of older personas were skewed toward viewing. The reading-centered class also leaned female (*OR* = 1.37), and the listening- and reading-centered classes concentrated in a few provinces, Sejong among them; however, every sex difference in average predicted probability was below 1 percentage point, and the largest provincial difference in the listening-centered probability was 5.0 percentage points. These differences show that particular content use contexts are coupled with particular demographic conditions in the NPK generation process.

## Implications for Digital Content Research

The present study shares with media repertoire research on actual user data (Lee & Park, 2020; Taneja et al., 2012) a focus on the configuration of engagement modes rather than on individual media. Nevertheless, types measured from actual use experience and types of mention in generated narratives are not the same object of study. For example, Lee and Park (2020) reported four types among actual news users, of which a television-dominant type (53.7%) was the largest and a multi-media consumption group that used several media together (8.9%) was the smallest. NPK resembles this in having a large class dominated by a single medium or mode, but because only 9.0% of NPK records mentioned two or more modes, no class corresponding to the multi-media type emerged. Whereas the Korea Media Panel Survey reported OTT and SNS use rates of 89.2% and 60.7%, respectively (Y. H. Kim, 2025), in NPK the viewing mention rate was 41.5% and the discovering-and-sharing mention rate was 7.0%. Both sources agree that video-related use is the most prominent, but whereas the SNS use rate in the actual survey was about two-thirds of the OTT use rate, discovering-and-sharing mentions in the NPK narratives were only about one-sixth of viewing mentions. Because a survey that asks about use experience and a coding scheme that captures mentions in free-text narratives measure different things, this numerical gap cannot be read as an exaggeration or understatement of actual behavior. It is nonetheless clear that, in the synthetic narratives, discovering-and-sharing contexts are generated in a contracted form relative to viewing contexts, which underpins the guideline in the following section that the narrative composition of synthetic data should not be read as the composition of the actual market.

The first contribution of this study is an analytic framework that classifies the content narratives of AI-generated synthetic personas by combinations of content engagement behavior rather than by the frequency of individual services. The Korean coding dictionary implementing this framework has been released together with its rules for handling ambiguous expressions, so that it can be applied to other synthetic data sets and in subsequent analyses.

The second contribution is a quantification of how the structured demographic attributes of synthetic personas are coupled with the generated content narratives. The three-step approach first fixes the latent classes and then shows which use contexts are generated more often under which demographic conditions; this implies that content research using synthetic data should examine not only the structured attributes but also how the generated narratives are conditioned on them.

**Implications for Content Planning and the Use of Synthetic Personas**

In content planning, the platform names and content tastes appearing in synthetic personas must not be interpreted as the observed behavior of actual consumers. For example, if synthetic personas of a particular age concentrate in a particular use-context type, this may reflect the actual use behavior of that age group, but it may equally be the generative model repeating a typified narrative it has learned.

On this premise, this study offers three practical guidelines for the use of synthetic personas. First, check whether the target use contexts of a content service correspond to the latent class composition of NPK. Second, if a particular class is heavily concentrated in certain ages, sexes, or regions, do not take that concentration as evidence of a real market segment. Third, in user simulation, select personas from different latent classes rather than relying on a single representative persona, so that the range of content responses is examined. Used in this way, the latent classes serve as an analytic tool for exploring the AI-generated narrative space and for detecting recurring content contexts and potential stereotypes. In particular, the finding that non-viewing contexts are rare among older personas identifies a concrete point at which to check the diversity of content narratives during generation and review.

**Limitations**

The limitations of this study are as follows. First, the five digital content engagement modes are an analytic classification constructed from the media repertoire perspective and the criterion of primary content engagement behavior; they are not a universal taxonomy applicable to the digital content industry as a whole. The precision audit relied on the judgment of a single author; no separate gold-standard data set was constructed, and recall was not estimated. Moreover, if new services, spelling variants, or generic expressions go uncaptured at different rates across groups, the contrast between the low-mention class and the other classes could be affected as well.

Second, explicit mention does not indicate actual frequency of use, preference, or satisfaction, and the absence of a mention does not indicate the absence of use. In addition, differences in the composition of mentions across age bands were not disentangled from the influence of narrative length or topic composition.

Third, because five binary indicators provide at most 32 response patterns and only 9.0% of records mentioned two or more modes, the latent classes are narrative types summarizing co-occurrence rather than broad use repertoires, and they do not guarantee that discrete groups exist. Furthermore, the three-step approach assumes that within-class mention probabilities are constant across age; given the discrepancies for some indicators noted in the Results (Table 6), the magnitudes of the age odds ratios are summaries conditional on this measurement structure. Models that allow direct effects of age on the indicators would change the definition of the classes themselves and were therefore beyond the scope of this study.

Fourth, because this study is limited to NPK version 1.0 and the specific system that generated it, the class structure and demographic associations may differ in other national editions, other generative models, or subsequent versions.

## Conclusion

This study classified the co-occurrence of viewing, discovering and sharing, interacting, reading, and listening in the 1,000,000 NPK records into latent classes and analyzed their associations with age, sex, and province of residence using a three-step approach, identifying four use-context types (viewing-centered, reading-centered, listening-centered, and low-mention), a steep shift in class membership with age, and a concentration of the listening and reading types in certain provinces.

These results show that AI-generated synthetic personas repeatedly generate particular digital content use contexts together with the demographic profile, and that the most pronounced pattern is the skewing of older personas' content mentions toward viewing. The types identified, and the demographic conditioning they reveal, provide an empirical criterion for inspecting the composition of synthetic persona data in content planning and user simulation. Using synthetic personas well requires looking beyond the plausibility of individual narratives to the content use contexts that recur across the data set and the demographic conditioning behind them; the typology developed here provides a tool for that inspection.

## Declarations

Data, materials, and code are available in Harvard Dataverse (https://doi.org/10.7910/DVN/SANW7A). No funding was received for this study. The authors declare no competing interests. During preparation of this work, the authors used DeepL to translate the initial draft and improve English readability, and OpenAI Codex to assist with refinement and error checking of the research code. The authors reviewed and edited all outputs and take full responsibility for the final manuscript and research code.

## Appendix

### Predicted Class-Membership Probabilities by Province of Residence

**Table A1**

*Predicted Class-Membership Probabilities by Province of Residence*

| Province | Viewing-Centered | Reading-Centered | Listening-Centered | Low-Mention |
|---|---|---|---|---|
| Seoul | 19.6 | 3.7 | 0.6 | 76.0 |
| Busan | 16.1 | 4.9 | 5.2 | 73.8 |
| Daegu | 18.6 | 5.1 | 5.0 | 71.3 |
| Incheon | 21.1 | 5.6 | 2.0 | 71.3 |
| Gwangju | 16.4 | 3.2 | 4.1 | 76.2 |
| Daejeon | 18.2 | 3.9 | 2.8 | 75.1 |
| Ulsan | 20.0 | 3.0 | 3.8 | 73.3 |
| Sejong | 18.3 | 8.3 | 5.4 | 68.0 |
| Gyeonggi | 22.9 | 3.4 | 1.9 | 71.8 |
| Gangwon | 18.9 | 3.0 | 0.4 | 77.7 |
| Chungbuk | 17.8 | 5.0 | 1.2 | 76.0 |
| Chungnam | 17.2 | 3.5 | 2.1 | 77.2 |
| Jeonbuk | 15.5 | 2.7 | 3.1 | 78.7 |
| Jeonnam | 16.6 | 2.6 | 3.0 | 77.8 |
| Gyeongbuk | 18.4 | 3.8 | 1.7 | 76.1 |
| Gyeongnam | 20.3 | 5.1 | 2.8 | 71.8 |
| Jeju | 11.9 | 3.5 | 3.9 | 80.6 |

*Note.* Values are class-membership probabilities (%) averaged over the sample distribution of age and sex. Seoul is the reference province; the full set of province coefficients is provided in the open materials.

**Precision Audit**

**Table A2**

*Match-Level Precision Audit by Engagement Mode*

| Final emitted mode | Correct/judged | Population matches | Weight | Precision | 95% interval |
|---|---|---|---|---|---|
| Viewing | 99/100 | 760,662 | .658 | .990 | [.946, 1.000] |
| Discovering & Sharing | 92/100 | 130,732 | .113 | .920 | [.848, .965] |
| Interacting (Gaming) | 100/100 | 109,937 | .095 | 1.000 | [.964, 1.000] |
| Reading | 100/100 | 88,790 | .077 | 1.000 | [.964, 1.000] |
| Listening | 99/100 | 65,348 | .057 | .990 | [.946, 1.000] |
| Weighted total | 490/500 | 1,155,469 | 1.000 | .984 | [.958, .991] |

*Note.* Units are final emitted match–mode pairs; one record may contribute multiple pairs. One author made all judgments. Per-mode 95% intervals are Clopper-Pearson exact intervals; the weighted interval is the central 95% interval of population-weighted precision simulated from independent beta posteriors with Jeffreys priors (Brown et al., 2001). The unweighted pooled precision was .980 (490/500); recall was not estimated.